\documentclass[english]{article}
\usepackage{graphicx,psfrag,color}
\def\bc{\begin{center}}
\def\ec{\end{center}}

\def\beq{\begin{equation}}
\def\eeq{\end{equation}}

\def\br{{\bf r}}

\def\bk{{\bf k}}

\def\br{{\bf r}}

\begin{document}

\title{
On the stability of the Mott phase in excitonic double layers
}

\author{Klaus Ziegler$^{1,2}$ and Roman Ya. Kezerashvili$^{2,3,*}$\\
%\affiliation{
%\address{
$^1$Institut f\"ur Physik, Universit\"at Augsburg, D-86135 Augsburg, Germany\\
$^{2}$Physics Department, New York City College of Technology,\\
 The City University of New York, Brooklyn, NY 11201, USA\\
$^{3}$The Graduate School and University Center,\\
The City University of New York, New York, NY 10016, USA\\
$^*$ email address of the corresponding author: rkezerashvili@citytech.cuny.edu
}
\date{\today}

\maketitle

\begin{abstract}
We study the stability of excitonic Mott phases in the presence of a periodic potential
and a nonlocal exciton-exciton interaction. 
The nonlocal interaction represents a classical Ising model and is treated in a 
mean-field approximation, while the local
repulsion of the excitons and the exciton tunneling are treated in a hopping expansion.
The convergence of the latter is the criterion for the stability of the Mott phase with respect
to quantum and thermal fluctuations.
This hybrid approach enables us to establish a stable bosonic Mott phase.
Our results could be useful to interpret recent experiments on electron-hole crystals in
van der Waals heterostructures.
\vspace{0.1cm}

\end{abstract}

{\it Introduction:}
The formation of an exciton gas in two-dimensional materials, as proposed and discussed by Lozovik
and Yudson~\cite{lozovik76}, has been an active research field for several decades.  
More recently, the formation of special ground states in interacting 2D electronic systems has 
attracted considerable attention since the discovery of graphene in 2005~\cite{Novoselov2005}. 
Besides magnetic ordering, there have also been reports on charge ordering and excitonic pairing 
with Mott phases in recent experiments~\cite{Geim2013,Qiu2024,Zheng2024}.
Motivated by experiments on $\alpha$-RuCl$_3$~\cite{Qiu2024,Zheng2024}
and on WSe$_2$/WS$_2$ layers~\cite{Alexeev2019,Fortin-Deschenes2024}, 
we analyze the conditions for the existence of a Mott phase in a double layer of excitons and
its stability in terms of quantum and thermal fluctuations.

\begin{figure}[h]
\begin{center}
\includegraphics[width=6cm]{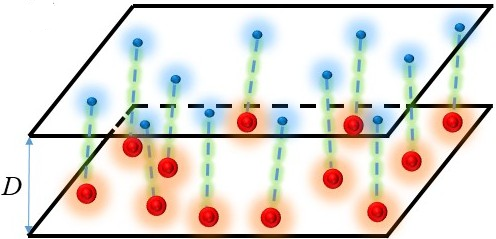}
\caption{
Schematic of the formation of an electron-hole gas in two parallel
layers separated by distance $D$, where the electrons reside in one,
the holes in the other layer. Due to the electron-hole Coulomb attraction, indicated by the
green lines, the electrons and holes form indirect excitons. The stability of the system requires 
an energy minimum. However, quantum fluctuations tend to destroy a spatial ordering.
}
\label{fig:RuCl3}
\end{center}
\end{figure}

One of the core tasks for the investigation of Mott insulators should address the rigidity of
the particle distribution. This depends strongly on the size of the gap relative to 
the hopping rate, as well as, to a lesser extend though, on the chemical doping of the material.
It is very challenging to tune the material properties by chemical doping the Mott insulator. 
A schematic picture of the excitonic Mott insulator is presented in Fig. \ref{fig:RuCl3}.
The electrons of the top layer and the holes in the bottom layer 
are subject to an attractive interlayer interaction, which leads to the formation of
excitons~\cite{lozovik76}. In this set-up the density of excitons is controlled through 
an external electric gate~\cite{Qiu2024}. Thus, the gate voltage induces a chemical potential 
$\mu$ for the exciton gas. 

The intralayer repulsion of electrons and of holes, respectively, can create
charge-density waves or generalized Wigner crystals~\cite{Regan2020}. Charge ordering can also be
caused by electron-phonon interaction~\cite{PhysRevB.105.085111}. 
Recent calculations, based on a self-consistent Hartree-Fock or a coherent potential 
approximation~\cite{PhysRevB.108.245113},
have revealed that a complex self-energy yields a reasonable approximation to describe experimental results, 
provided that the self-energy is calculated for the specific details of the material.
Without referring to a specific microscopic mechanism, an effective model for the description of an
excitonic gas in a double layer must include a mean-field approximation (MFA) for the exciton density
and the hard-core (HC) repulsion of the excitons due to the Pauli principle. 
The latter is crucial for
the description of quantum tunneling in the excitonic gas.
Using such a model, a central question is the stability of an ordered state of a 2D system in 
terms of quantum fluctuations.
For a Mott state, the phase fluctuations due to quantum tunneling are responsible for its instability. 
In other words, we can assume an ordered Mott state on a lattice in MFA and analyze 
its stability under an increasing rate of quantum tunneling. 
This approximation provides a phase diagram which only depends on the lattice structure as well as on the 
effective chemical potential and the hopping parameter, both scaled by the temperature.

{\it Model:}
For the following we assume that the electron-hole binding energy of the excitons exceeds all other energy 
scales in our model, such as the tunneling energy, implying that the excitons cannot dissociate.
This assumption is justified by a strong Coulomb interaction~\cite{BKZ2013}.
Moreover, the time scale of the observation is shorter than the recombination time such that the
electron and holes cannot transform into photons by recombination. 
The implementation of a lattice structure in the layers 
enables the excitons to form a commensurate Mott state by filling each well of the lattice with an exciton. 
A lattice can be realized either by the host material or by an electrically charged gate which is periodically 
structured \cite{BKLZPRB2019}.  The gate is used to control the density of excitons 
via its chemical potential $\mu$. Then a pure excitonic gas in a periodic potential $V_\br$
can be described by a hard-core Bose (HCB) gas with the many-body Hamiltonian
\beq
\label{eff_hamiltonian00}
H_{\rm HCB}
=-\sum_{\br,\br'\in\Lambda}J_{\br\br'}a^\dagger_{\br}a_{\br'}+H_I+h.c.
\eeq
with the density operator ${\hat n}_\br=a^\dagger_\br a^{}_\br$
\beq
H_I=-\sum_\br(\mu_0-V_\br){\hat n}_{\br}
+\sum_{\br,\br'}U_{\br-\br'}{\hat n}_\br {\hat n}_{\br'}
,
\eeq
where the sites $\Lambda=\{\br\}$ are the minima of the atomic material structure
and $J_{\br\br'}$ is a nearest neighbor tunneling (or hopping) rate:
\beq
J_{\br\br'}=\cases{J, & if $\br,\br'$ are nearest neighbors, \cr
0 & otherwise. \cr
}
\eeq
The lattice structure is characterized by the number of the nearest neighbors, which is 
the connectivity $c$. It represents the underlying atomic structure of the material.
There is a periodic potential $V_\br$, since the lattice can be modulated either by doping with 
other atoms/molecules or by strain, which are both described by the additional periodic superlattice.
Typical materials with this modulated lattice structure are transition metal
dichalcogenides (TMDCs), graphene/h-BN heterostructures or twisted 
bilayers~\cite{Zheng2024,Alexeev2019,Fortin-Deschenes2024,Regan2020}.
Finally, there is a translation invariant repulsive interaction $U_{\br-\br'}$, reflecting the weak
dipole-dipole interaction between the excitons.

The exciton creation operator $a^\dagger_r$ is composed of the electron creation operator 
$c^\dagger_{e,\br}$ and the hole creation operator $c^\dagger_{h,\br}$ as
$a^\dagger_\br=c^\dagger_{e,\br}c^\dagger_{h,\br}$. The form of the exciton operator $a^\dagger_\br$ 
implies that the excitons can tunnel freely with the hopping rate $J_{\br\br'}$ under the restriction that they
are composite particles which obey the hard-core condition $a^\dagger_\br a^\dagger_\br=0$
and the exciton commutator 
\beq
[a^{}_\br,a^\dagger_{\br'}]=\delta_{\br\br'}(1-c^\dagger_{e,\br}c^{}_{e,\br}
-c^\dagger_{h,\br}c^{}_{h,\br})
.
\eeq
The composite fermion statistics was used in the hopping expansion of 
Ref.~\cite{doi:10.1080/14786435.2016.1161860}. There we found that the difference of the
statistics between composite fermions and hard-core bosons does not affect  the estimation 
of the expansion terms. In other words, what is relevant is the fact that
at most one exciton can occupy a site of the lattice. Therefore, as a consequence of the 
Pauli principle of the fermions the exciton is approximately a HCB, similar to the 
exciton described by the effective Hamiltonian of 
Refs.~\cite{HANAMURA1977209,https://doi.org/10.1002/andp.20085200804}, and
a lattice system of strongly bound excitons is effectively a HCB gas. At
first glance, this excitonic gas has two complementary ground states, either a Bose-Einstein condensate (BEC) 
or a Mott state with at most one exciton per lattice site~\cite{https://doi.org/10.1002/andp.20085200804}. 
In the absence of a nonlocal exciton-exciton interaction and without an additional periodic single-particle potential
(i.e., $V_\br=0$), there is only a single Mott state with one exciton per site. On the other hand, 
with a repulsive nonlocal exciton-exciton interaction or a periodic single-particle potential $V_\br$, 
different Mott states are possible with commensurate fillings of the lattice. 

The BEC is described by a complex single-particle wavefunction 
$\Psi_\br=n_\br e^{i\varphi_\br}$, which is the solution of the Gross-Pitaevskii nonlinear
Schrödinger equation~\cite{gross61,pitaevskii61}. 
Here $n_\br$ is the exciton density and the phase $\varphi_\br$ of the wave function
is determined up to a
uniform phase shift $\varphi_\br+{\bar\varphi}$. The latter characterizes the BEC as a state
with global spontaneous symmetry breaking because a physical state must be unique. However,
the phase shift cannot be directly accessed experimentally such that there is no
need to fix it in the theory, for instance, by specific boundary conditions.
On the other hand, the Mott state with one particle per site is described by a many-body product state 
\beq
\label{Mott_state}
\Psi_{\br_1,\ldots,\br_N}=\prod_{j=1}^Ne^{i\varphi_{\br_j}}
.
\eeq
In contrast to the BEC, the phases $\{\varphi_{\br_j}\}$
are random with spatial correlations, which decay exponentially.
Other Mott states with partial lattice filling are product states with respect to the occupied lattice sites.
These complementary states should be approximate solutions of a linear Schrödinger equation 
for many particles with, for instance, the Hamiltonian (\ref{eff_hamiltonian00}), where the
Mott phase will be determined within the MFA.

We note that
there is a special effect of the HC properties of excitons, which distinguishes it
from a conventional Bose-Hubbard model.
Mott lobes occur for the bosonic Hubbard model with a finite local interaction
\cite{fisher89}.
In that case each lattice site can accommodate any
number of bosons, provided their chemical potential exceeds the finite
repulsive interaction. In the exciton model we assume a HC condition,
which can accommodate at most one boson per site. This fact prevents the formation
of Mott lobes~\cite{moseley08}. 
\begin{figure*}[t]
\begin{center}
\includegraphics[width=6.5cm]{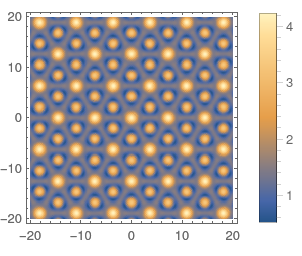}
\includegraphics[width=13cm]{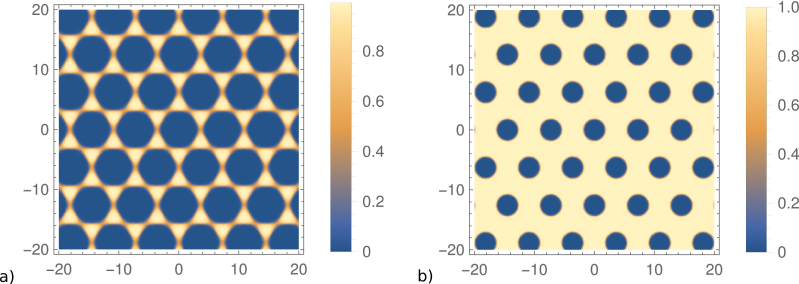}
\caption{The density $n_\br$ of the excitonic Mott insulator in the presence of a 
periodic potential with honeycomb structure 
$V_\br=\cos y+\cos(\sqrt{3}x/2+y/2)+\cos(\sqrt{3}x/2-y/2)$ (top panel): 
a) for $\beta\mu=-10$ and b) for $\beta\mu=10$.
%The periodic potential presents a honeycomb structure. 
$n_\br$ is calculated with Eq. (\ref{hc_mf2}) for $U_{\br-\br'}=0$. 
}
\label{fig:density}
\end{center}
\end{figure*}

{\it Excitonic Mott phase:}
In the absence of exciton hopping the HCB gas depends only on the
density operator ${\hat n}_\br$. Thus the eigenstates of the HCB Hamiltonian 
$H_I$ are Fock states $|\{n_\br\}\rangle=\prod_\br|n_\br\rangle$ with $n_\br=0,1$:
\beq
H_I|\{n_\br\}\rangle=
E_{\{n_\br\}}|\{n_\br\}\rangle
\ \ {\rm with}\ \
E_{\{n_\br\}}=-\sum_\br(\mu_0-V_\br) n_\br+\sum_{\br,\br'}U_{\br-\br'}n_\br n_{\br'}
,
\eeq
where the eigenvalues describe the energy of a classiscal Ising-like model with spin
$S_\br=n_\br-1/2$ in an external field.
Assuming a positive matrix $U_{\br-\br'}$, the ground state
for $\mu_0-V_\br\le 0$ is $|\{n_\br=0\}\rangle$, i.e., all sites of the lattice are empty. 
On the other hand, if $\mu_0-V_\br>0$ for some or all sites $\br$, the ground state 
may contain some occupied sites, representing a partially or completely filled lattice. 
The details of this ground state depend on the potential $V_\br$ as well as on the 
interaction $U_{\br-\br'}$.
A Fourier transformation $\br\to\bk$ diagonalizes $U_{\br-\br'}\to u_\bk$ such that
\beq
E_{\{n_\bk\}}=-\int_\bk[(\mu_0-v_\bk) n_\bk-u_{\bk}|n_\bk|^2]
.
\eeq
Then the ground state is $\prod_\bk|n_\bk\rangle$, where $n_\bk=0$ for all $\bk$
whose energy $\varepsilon_\bk=(\mu_0-v_\bk) n_\bk-u_{\bk}|n_\bk|^2$ is larger than 
the minimal energy $\varepsilon_{\bk_0}$. For instance, we get $\bk_0=0$ with
$n_{\bk_0}=1$ for 
a nearest-neighbor interaction $u_\bk=2-\cos k_x-\cos k_y$ and $v_\bk=0$.

When we consider thermal fluctuations, the knowledge of the ground-state properties
is insufficient and a thermal distribution must be considered, from which we can calculate
thermal expectation values. For instance, for the HCB gas without quantum
tunneling the average density reads
\beq
\label{hc_01}
\langle n_{\br}\rangle
=\frac{Tr\langle\{n_\br'\}|n_\br \exp(-\beta H_{\rm I})| \{n_\br'\}\rangle}
{Tr\langle\{n_\br'\}\exp(-\beta H_{\rm I})| \{n_\br'\}\rangle}
=\frac{\sum_{\{n_\br\}}n_\br\exp(-\beta E_{\{n_\br\}})}
{\sum_{\{n_\br\}}\exp(-\beta E_{\{n_\br\}})}
.
\eeq
For a complex interaction $U_{\br-\br'}$ it is well known that this can have different 
ordered phases~\cite{schulz81,villain81,uimin84}. 
In this case a MFA of the nonlocal interaction can be employed
\beq
\label{hc_mf0}
\sum_{\br,\br'}U_{\br-\br'}n_\br n_{\br'}
\approx \sum_{\br,\br'}U_{\br-\br'}n_\br \langle n_{\br'}\rangle
,
\eeq
where the average $\langle n_{\br}\rangle$ is taken with respect to the Boltzmann weight
$\exp(-\beta H_{\rm MFA})$ at inverse temperature $\beta=1/k_BT$, in which $H_{\rm MFA}$
is obtained from $H_I$ with the MFA (\ref{hc_mf0}). 
Then the thermal average of the density in Eq. (\ref{hc_01}) becomes the 
self-consistent relation
\beq
\label{hc_mf2}
\langle n_{\br}\rangle
=\frac{1}{1+\exp[-\beta(\mu_0-V_\br)]
\exp(\beta\sum_{\br'}U_{\br-\br'}\langle n_{\br'}\rangle)}
.
\eeq
It implies for $\beta\sim\infty$ a vanishing density $\langle n_{\br}\rangle\sim 0$ if 
$\mu_0-V_\br<0$ and a fully occupied lattice $\langle n_{\br}\rangle\sim 1$ if 
$\mu_0>V_\br + \sum_{\br'}U_{\br-\br'}\langle n_{\br'}\rangle$. 
The example in Fig. \ref{fig:density} for the honeycomb periodic potential 
$V_\br=\cos y+\cos(\sqrt{3}x/2+y/2)+\cos(\sqrt{3}x/2-y/2)$ and $U_{\br-\br'}=0$
illustrates that the density maxima appear in the potential minima and vice versa.
In the absence of any periodic potential $\langle n_\br\rangle$ is translation invariant
due to the translation invariance of $U_{\br-\br'}$:
\beq
\langle n\rangle=\frac{1}{1+\exp(-\beta\mu_0)\exp(\beta u_0\langle n\rangle)}
\ \ \
(u_0=\sum_{\br'}U_{\br-\br'})
,
\eeq
which gives $0\le \langle n\rangle \le 1$ as an increasing function of $\beta\mu_0$ and 
a decreasing function of $\beta u_0$.

The MFA in Eq. (\ref{hc_mf0}) can be understood as if there is 
an effective chemical potential 
$\mu_\br=\mu_0-V_\br-\sum_{\br'}U_{\br-\br'}\langle n_{\br'}\rangle$, which depends on the
solution of the self-consistent relation (\ref{hc_mf2}). Thus, there is an effective
HC Hamiltonian
\beq
\label{eff_hc}
H_{\rm HCB}\approx 
-\sum_{\br,\br'\in\Lambda}(J_{\br\br'}+\mu_\br\delta_{\br\br'})a_\br^\dagger a_{\br'}
+h.c.
,
\eeq
where only the HC (i.e. on-site)  exciton-exciton interaction
takes care of the repulsive interaction. We anticipate that this approximation is valid at
least for a dense exciton gas, which is relevant for the Mott state.
Then the HCB gas at inverse temperature $\beta$ can be further treated by employing a 
hopping expansion for a weak hopping rate $J>0$. 
This expansion starts from the free energy of the HCB gas $F=-\frac{1}{\beta}\log[{\rm Tr} e^{-\beta H}]$,
where the trace ${\rm Tr}$ is performed for the Fock basis with $n_\br=0,1$. Without hopping the free
energy reads $F_0=-\log[(1+e^{\beta{\hat\mu}})^{N/\beta}]$, where $N$ is the number of lattice sites, from 
which we get for low temperatures $F_0\sim -N{\hat\mu}\Theta(\beta{\hat\mu})$
with the Heavyside function $\Theta$. The average exciton density $\langle n\rangle$ becomes in this case 
\beq
\langle n\rangle=-\frac{1}{N}\frac{\partial F}{\partial {\hat\mu}}\sim\Theta(\beta{\hat\mu})
,
\eeq
which indicates an empty lattice for $\beta{\hat\mu}<0$ and a fully occupied  lattice with $\langle n\rangle=1$ for
$\beta{\hat\mu}>0$. This behavior represents that of an ideal Mott insulator for an HCB gas.

\begin{figure*}[t]
\begin{center}
\includegraphics[width=7cm]{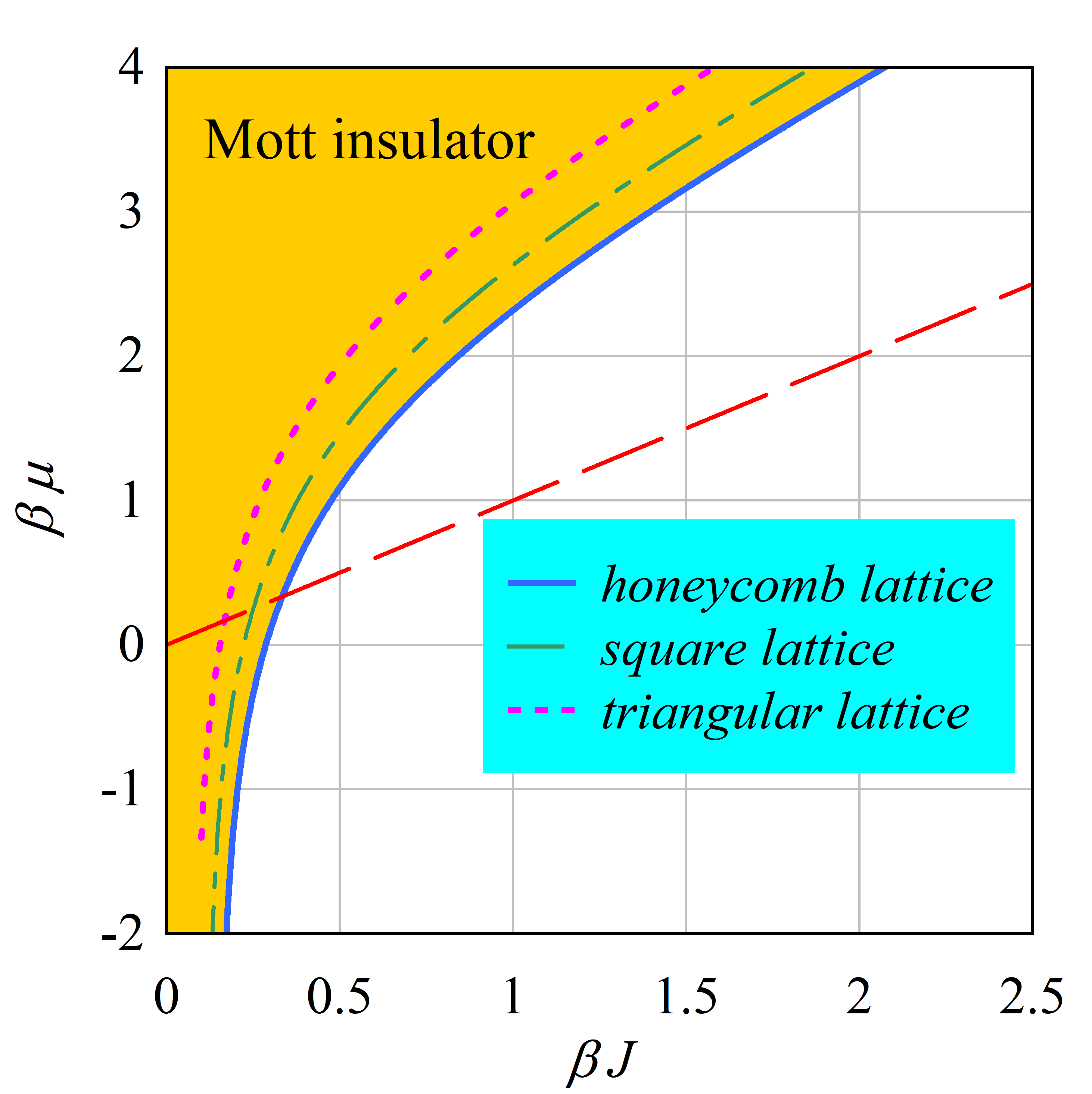}
\caption{Phase diagram from the hopping expansion of the HCB gas. 
The Mott insulator exists above the phase boundaries for the honeycomb lattice,
the square lattice and the triangular lattice, respectively,  which obeys the inequality 
(\ref{condition2}). The dimensionless chemical potential $\beta{\hat\mu}$ on the vertical axis
is obtained as ${\hat\mu}=\min_\br(\mu_\br)$.
The straight dashed red line is the phase boundary when the kinetic energy versus 
interaction energy obey the condition $\beta J<\beta{\hat\mu}$ for the Mott phase.
}
\label{fig:phase_diagram}
\end{center}
\end{figure*}

Now we analyze how this result is modified by a small hopping parameter $J$.
Following the functional-integral approach of Ref.~\cite{doi:10.1080/14786435.2016.1161860}
it turns out that the hopping expansion of $F$ separates into two expansions: (i) the expansion of the
exponential operator $e^{\beta H}$ and (ii) the expansion of the logarithm. The expansion series can
be estimated by a double sum of classical random walks, whose jump rates are estimated by
\beq
\Delta=\frac{c\beta Je^{\beta J}}{1+e^{\beta{\hat\mu}}-c\beta J}
.
\eeq
Thus, the hopping expansion converges for $\Delta<1$, which implies the condition
\beq
\label{condition2}
\beta cJ(1+e^{\beta J})<1+e^{\beta{\hat\mu}}
.
\eeq
In other words, the free energy for $J=0$ is only slightly disturbed by small hopping, in which the
probability of an exciton to hop away from its original site decays exponentially with the distance.
The criterion (\ref{condition2}) for the convergence was obtained after replacing $\mu_\br$ in the 
expansion by the effective chemical potential ${\hat\mu}=\min_\br(\mu_\br)$ of the MFA. 
Consequently, the stability of the excitonic Mott phase
depends on three parameters: the hopping rate $\beta J$, the lattice connectivity $c$ 
and the effective chemical potential $\beta\hat{\mu}$, scaled with the temperature.
For a given $c$, which is, for instance, $c=4$ for the square lattice, $c=6$ for the 
triangular lattice and $c=3$ for the honeycomb lattice, stable excitonic Mott phases are 
visualized in  Fig. \ref{fig:phase_diagram} by the yellow areas above the phase boundary 
$\beta cJ(1+e^{\beta J})=1+e^{\beta{\hat\mu}}$. 
At very low temperatures, which corresponds to large $\beta$, 
this phase boundary becomes the straight dashed line if $\beta{\hat\mu}=\beta J\gg 1$.
This reflects the simple energy argument that in the Mott phase the effective 
chemical potential dominates the kinetic energy. Our approach does not explain the
phase below the phase boundary, where the hopping expansion does not converge.
This phase may be a superfluid or an electron-hole plasma, both options have been
discussed in the literature, for instance, within a MFA~\cite{BKLZPRB2019}.
In contrast to the Mott phase though, there has not been a rigorous approach to this phase.

{\it Conclusions:}
We conclude that the stability of the excitonic Mott phase can be analyzed within 
a combination of a MFA for the nonlocal exciton-exciton interaction and 
a convergent perturbative expansion in terms of the exciton hopping. The latter 
provides an estimation of a stable Mott phase with a phase diagram.
It would be interesting to develop an analog
hybrid approach to study the regime where the hopping expansion does not
converge. This could provide an access to an excitonic superfluid and other 
collective phases. 
Another open problem, based on the approach of this work, is the analysis of the Mott 
phase within the Ising-like model for specific
combinations of a periodic potential and a translation invariant nonlocal
interaction, which can create complex excitonic density profiles.

%\bibliographystyle{unsrt}
%\bibliography{wigner_cryst}

\end{document}